\documentstyle[prl,aps,twocol]{revtex}
\begin{document}

\input psfig
\draft

\title{Parity-Affected Superconductivity in Ultrasmall Metallic Grains}

\author{Jan von Delft$^a$, Dmitrii S. Golubev$^b$, Wolfgang Tichy$^a$, 
Andrei D. Zaikin$^{a,b}$}

\address{$^a$
Institut f\"ur Theoretische Festk\"orperphysik, Universit\"at
Karlsruhe, 76128 Karlsruhe, Germany\\                               
$^b$P.N.Lebedev Physics Institute, Leninskii prospect 53, 117924,
Moscow, Russia }

\date{\today}

\maketitle

\begin{abstract}
We investigate the  breakdown of BCS superconductivity
in {\em ultra}\/small metallic grains as a function of particle size
(characterized by the mean spacing
$d$ between discrete electronic eigenstates), and
the parity ($P$ = even/odd) of the number of electrons on the island.
Assuming  equally spaced levels, we solve the parity-dependent BCS gap
equation for the order parameter $\Delta_P (d,T)$.
Both the $T=0$ critical
level spacing $d_{c,P}  $  and the critical
temperature $T_{c,P} (d)$   at which $\Delta_P = 0$ 
are parity dependent,  and both are so much 
smaller in the odd than the even case that these differences
should be measurable in current experiments.
\end{abstract}

\twocol
\narrowtext
The study of the properties of ultrasmall metallic
particles has witnessed a dramatic development during the last
year: using an ingenious new fabrication technique,
Black,  Ralph and Tinkham (BRT)  
\cite{RBT95}  have constructed a single-electron
transistor (SET) whose island, a single nm-scale Al grain,
is more than four orders of magnitude smaller in volume
(estimated radii between $r \sim 2.5$ and 13~nm) than that of
conventional SETs. Thus a new energy scale,
the average level spacing $d = 1/N( \varepsilon_F)$
between discrete electronic levels, enters the problem:
Both the free-electron estimate of 
$d \simeq 2 \pi^2 \hbar^2 /(m k_F
{\cal V})$ and direct observation  (discrete steps in the $I$-$V$ curve)
give values of $d$ ranging from 0.02 to 0.3 meV, the latter being
much larger than the smallest accessible temperatures
$(\simeq 30$~mK) and on the order of the bulk superconducting
gap ($\Delta_b = 0.18$~meV for Al).

The eigenenergies of the larger grains ($r > 5$ nm) studied by
BRT revealed the existence of an excitation gap $\Omega > d$
which is driven continuously to zero by an applied magnetic field, 
and striking gap-dependent 
parity effects, i.e.\ differences between islands
with an even or odd [$P=e/o$] number of electrons.
BRT very convincingly interpreted these phenomena 
as evidence for  superconductivity.
On the other hand, in smaller particles ($r \!<\! 5$ nm) 
it was not possible to detect an $H$-dependent gap
or other evidence for superconductivity.

These experiments invite reconsideration of an old 
but fundamental question:
{\em what is the lower size limit for the existence of superconductivity
in small grains?}\/
Anderson addressed this question already in 1959 \cite{And58} and 
argued that ``superconductivity would no longer be possible''
if the level spacing $d$ becomes larger
than the bulk gap $\Delta_b$, for reasons explained below.
This answer -- although in general correct -- is not yet quite 
complete, since it does not address {\em parity effects}\/.
 Even in  ``large'' 
superconducting islands (with $d \ll  \Delta_b$)
experiments \cite{HS} have demonstrated
the dramatic impact of parity on  $I$-$V$ characteristics;
moreover theory \cite{JSA94,GZ94} 
predicts an even-odd  difference for the
{\em superconducting order parameter
itself}\/  of  $\Delta_e \!-\! \Delta_o \!=\! d/2$ at $T=0$.
Though the \mbox{latter}  dif-
\vfill
\vglue6truecm 
\noindent
ference is immeasurably small in ``large'' 
islands, it  should
certainly become significant in ultrasmall grains.
Moreover, since the crossover temperature  at which parity 
effects  become  observable \cite{HS}, namely 
$T_{cr} =  \Delta_b /
\ln N_{eff}$ (where in the $d \ll  \Delta_b$ limit 
$N_{eff} = \sqrt{8 \pi T \Delta_b}/d$),
 becomes of order $ \Delta_b$ when $d \simeq
\Delta_b$, parity effects should survive to temperatures
as high as $T_c$ itself. Hence $T_{c,P} (d)$
as function of $d$ should be parity-dependent too.

In this Letter we address these issues by studying 
  parity effects in the order parameter $\Delta_P (d, T)$
for general $d$. In particular, we calculate
 $\Delta_P (d, 0)$ and 
 $T_{c,P} (d)$ by solving the BCS
gap equation (derived using parity-projected  mean-field theory 
\cite{JSA94,GZ94}) at $T\!=\!0$ and $\Delta_P \!=\! 0$, respectively, 
for the case of equally-spaced single-particle levels.
We find $T_{c,o} (d) / T_{c,e} (d) < 1$
and a remarkably  small ratio of 
critical level spacings $d_{c,o} / d_{c,e} \!=\! 1/4$  at $T\!=\!0 $.
Our results are completely compatible
with BRT's observations. Moreover,  the
predicted parity differences should be directly observable
in their latest  experiments which have variable gate voltage, 
allowing them to change the number parity of a given grain at will.

{\em The model:}
In BRT's experiments, the charging energy $E_C \!=\! e^2 /2 C_{total}$
of an ultrasmall grain is by far the largest energy scale in
the problem (with $E_C \simeq$ 4 meV $\gg \Delta_b$),
so that fluctuations in particle number are strongly suppressed.
Therefore, in this Letter we   consider a completely isolated grain,
which should be described using a canonical ensemble
with a prescribed  number of electrons $n \!=\! 2m + p$,
where $p \!=\! (0,1)$ for $P \!=\! (e, o)$ (the labels $p$,
$P$ {\em and also $n$}\/ will be used interchangeably as parity
labels below).
We adopt a model Hamiltonian having the standard reduced BCS form:
\begin{equation}
\label{Hamiltonian}
	\hat H =  \sum_{j \sigma} \varepsilon^0_{ j}
	c^\dagger_{j \sigma} c_{j \sigma}
 	-  \lambda d \sum_{ij}{}'
	c^\dagger_{i +} c^\dagger_{i -}
	c_{j -} c_{j +} \; .
\end{equation}
Here $c^\dagger_{j \sigma} $
creates an electron in the particle-in-a-box-like,
	independent-electron state $|j\sigma \rangle$,
where the states $|j + \rangle$ and $|j -\rangle$ 
are degenerate, time-reversed partners whose
energies $ \{ \varepsilon^0_{ j} \}$
are considered as a given set of phenomenological parameters.
The integer $j$ is a  discrete quantum number.
For a given $n\!=\!2m+p$, we take $j\!=\!0$ to describe the  
first energy level whose occupation in the $T\!=\!0$ Fermi sea
is not 2 but $p$, so that $j = -m , \dots \infty$.
Finally,  the dimensionless coupling constant  
$\lambda^{-1} = \ln {2 \omega_c \over \tilde \Delta}$ is regarded 
as a phenomenological parameter determined
by the  value $\tilde \Delta \equiv \Delta (0,0)$ of the 
effective gap (measured at  $d \ll \tilde  \Delta$) and
 some cut-off frequency $\omega_c$.

{\em Pair-mixing:}\/
At this point it seems appropriate to briefly address the question of
what is meant by the  ``existence of superconductivity'' in
ultrasmall grains. It deserves special attention,
firstly because the BCS order parameter
$\Delta$, if defined as usual
by $ \lambda \, d \sum_j \langle c_{j -} c_{ j + }
\rangle$, seems to be zero in a canonical ensemble,
and secondly because 
most of the standard criteria, 
e.g. a gap  followed by a continuous excitation spectrum,
zero resistivity and the Meissner effect, are not applicable here.

Now, the microscopic reason for all of these (large-sample) 
phenomena is of course 
the {\em existence of a pair-correlated ground state.}\/
The essence of its correlations is what we shall call
{\em pair-mixing}\/ across $\varepsilon_F$, namely
the partial population (with amplitude $v_j \!>\! 0$) 
of some time-reversed
pairs of states $(| j +\rangle, | j - \rangle )$
above $\varepsilon_F$ $(j \!>\! 0)$ by partially depopulating
(with amplitude $u_j \! > \! 0$) 
some pairs of states below $\varepsilon_F$ $(j \!<\! 0$).
This creates phase space 
for pair scattering (which is Pauli-blocked in
the normal ground state) and hence allows the BCS 
interaction to lower the ground state energy.

Although BCS showed that a brilliantly simple way of calculating
the $u_j$ and $v_j$ is to use grand-canonical methods, 
 pair-mixing of course can and does also occur in a fixed-$n$ system. 
Indeed, this pair-mixing can readily be characterized by
a ``generalized'' order parameter that  is equal to the conventional
 $ \lambda d \sum_j \langle c_{j -} c_{ j + } \rangle$
in \mbox{BCS's} 
grand-canonical mean-field treatment,  but (in contrast to
the latter expression) is meaningful in a fixed-$n$ system too,
namely $ \lambda\, d  \sum_j u_j v_j$.
An experimental signature of this pair-mixing
is the energy cost needed to  add or remove single
electrons that perturb  these correlations (i.e.\ that
``break pairs''). 
Since BRT quite unam\-bi\-guously measured
such energy costs in their larger grains,
it seems reasonable to regard
these  as ``superconducting'',  in the
sense of having a {\em pair-correlated ground state
 that measurably exhibits pair-mixing.}

The notion of pair-mixing also provides a simple
way to understand why superconductivity 
ceases to exist in sufficiently small samples.
If the level spacing \mbox{becomes} sufficiently large
($d \!\simeq \! \tilde \Delta$),
pair-mixing costs a \mbox{prohibit}\-ive
amount of kinetic energy and hence ceases to  occur.
The task at hand is to describe this breakdown
(semi)\-quantitatively,  while keeping track of parity effects.

{\em Canonical and Parity Projection:}\/
Since in practice it is so much harder
to calculate $u_j$, $v_j$ canonically than 
grand-canonically, and since the latter approach
has been remarkably successful
even in small systems such as nuclei
with as little as $n \sim 100$ nucleons \cite{Iachello}, 
purely canonical descriptions (pioneered by Schafroth \cite{Blatt})
are seldom used.
An alternative approach \cite{JSA94,GZ94} is to 
employ an auxiliary parity-projected grand-canonical
partition function, 
\begin{equation}
\label{ZGP}
	Z_P^G (\mu) \equiv  \mbox{Tr}^G 
	{\textstyle {1 \over 2}} [1 \pm (-)^{\hat N}]
	 e^{- \beta (\hat H  - \mu \hat N)} \; \equiv \; 
	e^{- \beta \Omega_P^G (\mu)} \; ,
\end{equation}
($\mbox{Tr}^G$ denotes a grand-canonical trace),
from which the desired  fixed-$n$
partition function $Z_n$ can in principle be exactly projected by 
integration:
\begin{equation}
\label{Zn}
	Z_n = \int_{- \pi}^\pi {du \over 2 \pi}
	e^{- i u n}
	Z^G_P (i u / \beta) \; .
\end{equation}
Since in practice, though,  it is hard to perform the integration
exactly,  we approximate the integral by its saddle point value,
$Z_n \! \simeq \!  e^{- \beta \mu_n n} Z_P^G (\mu_n)$,
where $\mu_n$ is fixed by
\begin{equation}
\label{mun}
	n = - 
	\partial_\mu \Omega_P^G (\mu) \big|_{\mu = \mu_n}
	\quad \bigl[ \;= \langle \hat N \rangle_P \; \bigr] \; 
\end{equation}
(Here $\langle \quad \rangle_P$ 
 is taken in the parity-projected grand-canonical ensemble of
$Z_P^G$.) 
This equation, the bracketed part  of which is 
the parity-projected version of a standard
grand-canonical identity, illustrates the elementary fact
that the saddle point approximation produces nothing but the
 grand-canonical description we had set out to improve upon.
Nevertheless, the merits of the above approach  are that
(i) corrections can be calculated systematically by including Gaussian
contributions [governed by 
$(\langle N^2 \rangle - \langle N \rangle^2)^{-1/2}$]
 to the integral~(\ref{Zn})
(at $T\!=\!0$ they are $\sim (d / \Delta)^{1/2}$
\cite{JSA94});
(ii) it illustrates that the parity projection of Eq.~(\ref{ZGP}),
which is essential for extracting $e/o$ differences,
can be done exactly even when the fixed-$n$ projection cannot;
(iii) it clarifies that in
a canonical ensemble $\mu_n$ is  simply the saddle-point value
of an integration parameter, which, however,
has to be determined with special care
in ultrasmall grains, for which $d$ is large. 

{\em Mean-Field Approximation:}\/
We evaluate $Z_P^G$ using a ``naive mean-field approach''
(our method is equivalent to that used in \cite{GZ94}):
Make the replacement 
\begin{equation}
	c_{j-} c_{j+} \to \{ c_{j-} c_{j+} - 
	\langle  c_{j-} c_{j+} \rangle_P \} 
	+ \langle  c_{j-} c_{j+} \rangle_P \; 
\end{equation}
in $\hat H \!-\! \mu_n \hat N$, 
neglect terms quadratic in the fluctuations
represented by $\{ \quad \}$ and diagonalize,
using $\gamma_{nj\sigma} = u_{nj} c_{nj\sigma} - \sigma v_{nj} c^\dagger_{nj
- \sigma}$. One obtains the usual  results
	$\hat H \!-\! \mu_n \hat N \simeq 
	C_n + \sum_{j \sigma} E_{n j \sigma} 
	\gamma_{n j \sigma}^\dagger \gamma_{n j \sigma}$,
where 
	$E_{n j \sigma} \!=\! [\varepsilon_{nj}^2 +
	\Delta_P^2]^{1/2}$ ,
	$\varepsilon_{nj} \!\equiv\! \varepsilon_{j}^0 - \mu_n$, 
and
$v_{nj}^2 \!=\! {1 \over 2}  (1 - \varepsilon_{nj} / E_{nj})$.
Moreover, since the parity of electron number and quasiparticle
number are always the same, Eq.~(\ref{ZGP}) can be rewritten
 \cite{JSA94} using  quasiparticle-parity projection,
$Z_P^G (\mu_n) = {1 \over 2} (Z^G_+ \pm Z^G_-)$, where 
\begin{equation}
\label{ZBGBCS}
      Z_\pm^G (\mu_n) = e^{-\beta C_n}
	\prod_{j \sigma} (1 \pm e^{- \beta E_{n j \sigma}}) \; .
\end{equation}
The usual mean-field self-consistency condition
$	\Delta_P = \lambda \, d \sum'_j
	 \langle c_{ j -} c_{ j +} \rangle_P
$
takes the form
\begin{equation}
\label{gap}
	{1 \over \lambda} = d  \sum_{|j| < \omega_c / d}
	{1 \over 2 E_{n j}} \left( 1 - \sum_\sigma f_{n j \sigma} \right) \; ,
\end{equation}
where 
$f_{nj\sigma} = \langle \gamma_{nj \sigma}^\dagger \gamma_{nj \sigma} 
\rangle_P $ is given by \cite{JSA94,GZ94}
\begin{equation}
\label{fnisigma}
	f_{nj \sigma} = 
	- {\partial \over \beta \partial_{E_{nj \sigma}}}
	 \ln Z_P^G (\mu_n) = 
	f_{n j \sigma}^+ + 
	{ f^-_{n j \sigma} -  f^+_{n j \sigma} \over 1 + (-)^p 	R} \; .
\end{equation}
Here we defined 
	$f^\pm_{nj \sigma} \equiv \pm (e^{\beta E_{nj \sigma}} \pm 1)^{-1}$
and 
	$R \equiv {Z_+^G \over Z_-^G}$.
The above description thus involves
 the usual BCS quasiparticles, but {\em their number
parity is restricted to be $p$}.

Corrections to the naive mean-field expressions
for $E_{nj\sigma}$  and $\Delta_P$
[as measured, e.g., by fluctuations in $\Delta_P$, or
by $( \lambda d  \partial_\Delta^2 \Omega_P^G)^{-1/2}$]
can be shown to be of order  $\lambda  d \,  / \Delta$.
Janko {\em et al.}\/ \cite{JSA94} calculated
the leading one of these corrections (their $1 / N_{eff}$ terms) by
 using a more careful
 mean-field approach which incorporates contributions
from the $\{ \quad \}^2 $ terms  neglected above.
However, for $\lambda d / \Delta$ sufficiently
large that these terms matter, other non-mean-field
corrections ($\simeq (\lambda d / \Delta)^{1/2}$) matter too.
Nevertheless, since our goal is merely to find a criterion 
for when pair-mixing correlations cease to exist,
and one such criterion is to establish when
the mean-field gap equation no longer has a solution, 
 we shall neglect all corrections going beyond 
naive mean-field theory.

{\em Determination of $\mu_n$:}\/ 
Following \cite{STKC70},
let us henceforth consider the case of equal level spacing,
$\varepsilon_{j}^0 = j \, d + \varepsilon^0_0$ (which seems reasonable 
for large $n$, due to level repulsion). 
Using $ 	\langle \hat N \rangle_P \!=\! \sum_{j \sigma}
	\left( v_{nj}^2 + (u_{nj}^2 - v_{nj}^2) f_{nj \sigma} \right)
$
and symmetrizing the $j$-summation interval 
by  dropping the negligibly small 
 $j \!>\! m\!-\! \delta_{P,e}$ terms, 
 Eq.~(\ref{mun}) gives \cite{JSA94}
$$
	2 m + p =\! \!\! \!\sum_{j = -m}^{m-\delta_{P,e}}  \!\!
	\left[2 \theta (\varepsilon_{nj}) +\!
	\left\{ \! \mbox{sign}(\varepsilon_{nj}) \! - \!
	{\varepsilon_{nj} \over E_{nj}} \! \right\}
	+ \! {\varepsilon_{nj} \over E_{nj}} 2 f_{nj} \right]  .
$$
Here we have added and subtracted 
$ \sum_j \mbox{sign}(\varepsilon_{nj})$
(defining  sign$(0) \equiv 0$ and $\theta (0) \equiv {1 \over 2}$).
By inspection, this implies that 
$\mu_n = \varepsilon^0_0 - {1 \over 2} d \, \delta_{P,e}$,
because then
$
	\varepsilon_{nj} = d (j + {\textstyle {1\over 2}}
	\delta_{P,e}) \; ,
$
which ensures firstly that
$2 \theta( \varepsilon_{n0}) = p$, and 
secondly that the summands of the second and third terms
are anti-symmetric, so that these terms vanish.
(More generally, they vanish also 
for non-equally-spaced levels as long as
the levels are distributed roughly symmetrically 
about $\varepsilon_0^0$, in which case
$\mu_n \simeq 
 \varepsilon_0^0 \delta_{P,o} +  {1 \over 2} 
( \varepsilon^0_{ 0} - \varepsilon^0_{ -1})  \delta_{P,e}  $.)
Thus we have proven the seemingly obvious:
in the language of the $T\!=\!0$ normal Fermi sea,
 $\mu_n$ lies
exactly half-way between the last filled and first empty
levels if $P\! = \! $ even, and exactly on the singly occupied level
if $P\!=\!$ odd.

We are now ready to study the gap equation  (\ref{gap}).

{\em Gap Equation at $T\!=\!0$:}\/
The quasiparticle occupation function 
reduces to $f_{nj \sigma} \!=\! {1 \over 2} \delta_{j 0} \delta_{P,o}$
at $T\!=\!0$, as intuitively expected, because then the even or odd systems
have  exactly zero or one quasiparticle, the latter in the lowest
quasiparticle state, namely $j\!=\!0$. This $e/o$ difference has
a strong impact on the $T\!=\!0$ gap equation:
in the odd case, the $j\!=\!0$ level, for which 
$E_{n j}^{-1}$ is largest, is {\em
absent}\/, reflecting the fact that the odd quasiparticle in the
$j\!=\!0$ state obstructs pair scattering involving this state. To
compensate this missing term, $\Delta_o$ must therefore become
significantly smaller than $\Delta_e$ as soon as $d$
is large enough that a single term becomes significant relative
to the complete sum.

To quantify this statement, it is convenient to
rewrite Eq.~(\ref{gap}) at $T\!=\!0$ as follows:
Writing $E_{nj}^{-1} \!=\!
 \int \! d \omega \! /\!   \pi\,  (E_{nj}^2 + \omega^2)^{-1}$,
transferring the cut-off  $\omega_c$
from $\sum_j$ to  $\int \! d \omega$, and performing
the $j$-sum (by contour integration) gives 
\vglue-0.1truecm
\noindent
\begin{equation}
\label{gapT0}
	\ln {2 \omega_c \over \tilde \Delta} = \!\!
	\int_{0}^{\omega_c} \! {d \omega  \over E_{P \omega}} \!
		\left[
	\left(\tanh{ \pi E_{P \omega} /  d}\right)^{1- 2p}
	\!\! - { d \, \delta_{P,o}  \over \pi
	E_{P \omega}} \right] \! ,
\end{equation}
\vglue-0.1truecm
\noindent
where $E_{P \omega} = (\omega^2 + \Delta_P^2)^{1/2}$.
Since, amusingly,  for $P=\!e$  Eq.~(\ref{gapT0}) is identical in form 
(with $d \to  2 \pi T$)
to the well-known  gap equation for the $T$-dependence of the 
bulk gap  (curve A in Fig.~\ref{fig:3D}), we have 
 $\Delta_e ( d, 0)= \Delta_P (0,  d/2 \pi) $. 
Thus, for $d / \tilde \Delta \ll 1$  the even gap has the
standard form 
$\Delta_e (d, 0) = \tilde \Delta (1 \! - \! \sqrt{\tilde \Delta d} \, 
e^{- 2 \pi \tilde \Delta /d})$; 
in contrast, one easily finds from Eq.~(\ref{gapT0}) that
  the odd gap drops linearly, 
$\Delta_o (d, 0) = \tilde \Delta  -  d/2$,
in agreement with \cite{JSA94,GZ94}.

The full solutions of Eq.~(\ref{gapT0}) for
$\Delta_P        (d, 0)$,  obtained numerically,
  are shown as curves B and C in Fig.~\ref{fig:3D}.
The critical values $d_{c,P}$ at which $\Delta_P (d_{c,P},0)= 0$
can be found analytically by setting $\Delta_P \!= \!T\! = \! 0$ in 
Eq.~(\ref{gap}):
\begin{equation}
\label{dcritical}
	{d_{c,e} \over \tilde \Delta}  = 2 e^\gamma \simeq 3.56 
	\qquad \!\! \mbox{and} \!\! \qquad
	{d_{c,o} \over \tilde \Delta} 
	 = {\textstyle {1 \over 2}} e^\gamma \simeq 0.890 \; .
\end{equation}

{\em Critical temperature:} Although  ultrasmall
grains cannot undergo a sharp thermodynamic
phase transition (this would require $n \to \infty$), 
the quantity $T_{c,P} (d)$,
defined simply as the solution to the $\Delta_P \to  0$ limit
of  Eq.~(\ref{gap}),
is another measure of how rapidly pair-mixing correlations
break down as function of level spacing.
Our numerical results for  $T_{c,P} (d)$ \cite{TvD},
shown as  curves D and E of Fig.~\ref{fig:3D} for $P=e/o$,
have the expected limits at $d = 0$ and $ d_{c,p}$, but 
behave unexpectedly in between: 

{\em Even:} In the even case, $T_{c,e} (d)$ is non-monotonic, 
increasing slowly from its initial value $ T_{c,e} (0)
\!=\! \tilde \Delta e^\gamma / \pi \!=\! 0.567  \tilde \Delta $  
to a maximum given by
$T_{c,e} (2.60 \tilde \Delta) \! = \!					 0.736 \tilde \Delta > T_{c,e} (0)$.
It then drops to zero very rapidly
as $T_{c,e} (d) \simeq d / \log[4 d_{c,e}/ (d_{c,e} - d)]$,
which follows from setting $\Delta_e = 0$
in
$
	\Delta_e (d, T) \simeq d \left[
	2 (1 - d/ d_{c,e} - 4 e^{-d/T}) /( 7 \zeta (3)) \right]^{1/2} 
$,
an analytical result that holds for 
 the regime $T, \Delta_e (T)  \ll d$. 
The intuitive reason for the initial increase in $T_{c,e} (d)$ is
that the difference between the 
actual and the usual quasiparticle occupation
functions is $f_{nj\sigma} \!-\! f_{nj\sigma}^+ \!<\! 0$ 
for an even grain (becoming significant when $d \simeq \tilde \Delta$), 
reflecting the fact that exciting quasiparticles two at a time
is more difficult than one at a time. Therefore
 the quasiparticle-induced break-down of superconductivity with
increasing $T$ will set in at slightly higher $T$
if $d \simeq \tilde \Delta$.

{\em Odd:} In the odd case, $T_{c,o} (d)$ ``overshoots'' a bit
near $d = d_{c,o}$, or alternatively, the critical level spacing
$d_{c,o} (T)$ is  non-monotonic as a function of increasing $T$,
reaching a maximum value given by 
$d_{c,o} (0.387 \tilde \Delta) = 1.05 \tilde \Delta$
before beginning to decrease. The intuitive reason
for this is that for $0 < \Delta_o \ll T,d$
the odd $j=0$ function $f_{n0\sigma} (T)$
becomes somewhat smaller than its $T=0$ value of ${1 \over 2}$,
because with increasing $T$  some of the probability
for finding a quasiparticle in state $j$ ``leaks'' from
$j=0$ to higher states with $j \neq 0$, for which
$E^{-1}_{nj} < E^{-1}_{no}$ in Eq.~(\ref{gap}). Thus, the
dramatic blocking-of-pair-scattering effect of
the odd quasiparticle becomes slightly less dramatic as $T$
is increased, so that  $d_{c,o}$ increases slightly.

{\em Discussion:}\/ 
An important general feature of our results
is  that level discreteness {\em always reduces}
$\Delta_P(d,0)$  to be  $<  \tilde \Delta$
(thus contradicting Ref.~\cite{Parmenter}, which was convincingly
criticised in Ref.~\cite{STKC70}). However,
BRT's experiment found an effective gap $\tilde \Delta$ that is larger
by a factor of 1.5 to 2 than its bulk value $\Delta_b$.
 Following the argumentation of \cite{STKC70} for thin films,
we can attribute this to presumed changes in the phonon spectrum in  small
samples,  which can be modeled by using a constant
value of $\lambda$ 
larger (by a few percent) than the usual bulk 
value $\lambda_b$. 

The rather rapid drop of $\Delta_P (d)$, once it happens,
could be the reason why BRT see a well-developed gap $\tilde \Delta$
even for $d \simeq \tilde \Delta$, but don't see any for
their smallest grains.
 The most important conclusion of this paper, though, 
is summarized by Fig.~\ref{fig:3D} and Eq.~(\ref{dcritical}):
there is a large regime in which  $\Delta_o = 0$ while
 $\Delta_e$ is still $\simeq \tilde \Delta$, in other words,
{\em pair-mixing correlations vanish significantly sooner
 for odd than even grains as their size is reduced}\/.
Moreover, the  surprising largeness of the
  ratio $d_{c,e} / d_{c,o} = 4$ 
opens the exciting possibility of studying
grains with  $d_{c,o} < d < d_{c,e}$, which should have
 $\Delta_o = 0$ while $\Delta_e$ is still $\simeq \tilde
\Delta$. Since this leaves tell-tale traces in
the $I$-$V$-$V_g$ characteristics \cite{TvD},
 BRT should be able to test this prediction directly, 
since they can change the electron parity of a given 
grain in a controlled way by tuning the gate voltage of
their SET, and hence  study 
the {\em same}\/ grain in both its even and odd states.

Of course, a nagging question remains:
How robust are our mean-field-based results?
Will not fluctuations in $\Delta$,  governed by
$d  /\tilde \Delta$ and hence large in ultrasmall
grains, wash out the predicted 
parity differences in  $\Delta_{e/o}$?
For instance, it is doubtful that
the  unexpected non-monotonic subtleties  of $T_{c,P} (d)$,
though intuitively plausible, have physical
significance, since they fall in the
$\Delta_P \simeq 0 $ regime where fluctuations
are extremely large.

However, while a detailed analysis of the fluctuations
is beyond the scope of this Letter, we believe that,
 at least in the (experimentally accessible) regime of 
$T/d \simeq 0$,
our main result is indeed robust: Firstly, it
is known that at $T=0$, the leading $d / \tilde \Delta$ 
fluctuation corrections to the free energy difference
$\Omega^G_e - \Omega^G_o$ are parity-independent
\cite{JSA94,GZ94}. Secondly, since as $d$ approaches
the point where
$\Delta_o$ vanishes, 
we have $ d/ \Delta_e \lesssim 1 \ll d / \Delta_o $, 
fluctuations become important much later
in  the even than the odd case.
Thirdly, even for systems much
smaller than ultrasmall grains (that have $n \sim 10^4$),
namely shell model nuclei (with  $n \sim 100$),
the $T=0$ BCS-description of pairing interactions 
has been remarkably successful \cite{Iachello},
despite the presence of large fluctuations.

The perhaps most compelling support for 
the reality of the predicted effect comes from 
the following back-of-the-envelope
argument: the ground state energy difference between a BCS
superconductor with parity $P$ 
and the normal  Fermi sea is 
$- {\tilde \Delta_P^2 / ( 2 d)} + \tilde \Delta_o \delta_{P,o}$. 
While for $P\!=\!e$ it is always negative, for $P\!=\!o$  it vanishes
when $d \!> \!{1\over 2} \tilde \Delta$,  illustrating that
the odd electron  indeed dramatically
disrupts pair-mixing correlations when $d \! \simeq\! \tilde \Delta$.

In conclusion, we have investigated the influence of parity 
on the superconducting mean-field order parameter in ultrasmall
grains. We have found that as a function of decreasing
grain size, superconductivity breaks down 
in an odd grain significantly earlier than in an even grain,
which should be observable in present experiments.

It is a pleasure to thank BRT for showing us their
 preliminary results,
and to acknowledge  discussions with
V. Ambegaokar, 
C. Bruder, B. Janko, 
H. Kroha, A. Rosch, G. Sch\"on and J. Siewert.
This research was supported by ``Sonderforschungsbe\-\mbox{reich}
 195'' of the Deutsche \mbox{Forschungsgemeinschaft}{\vspace{-2cm}.

\begin{figure}
\centerline{
\psfig{figure=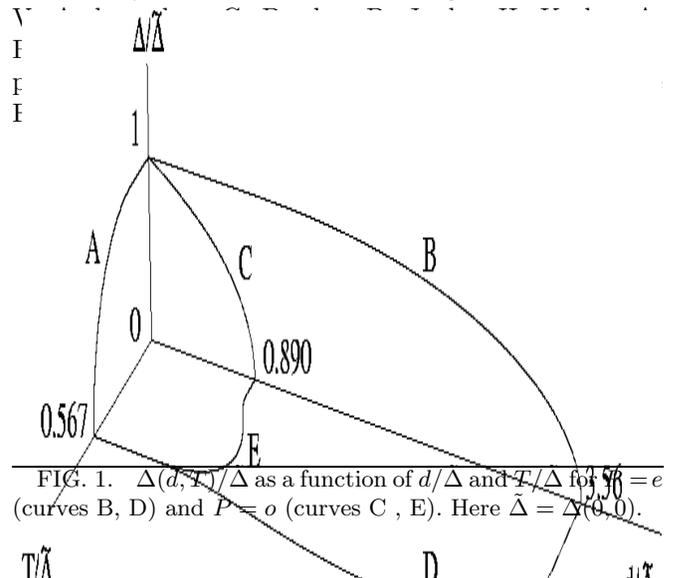,height=8cm,width=8.5cm}}
\vspace{-2cm}
\caption{\label{fig:T=0} $\Delta(d, T) / \tilde \Delta$
as a function of $d/\tilde \Delta$ and $T/\tilde \Delta$
for $P=\!e$ (curves B, D) and $P=o$ (curves C , E).
Here $\tilde \Delta = \Delta (0,0)$.
\label{fig:3D}}
\vspace{-2cm}
\end{figure}

\end{document}